\documentclass[aps,prb,twocolumn,groupedaddress,showpacs]{revtex4}

\usepackage{graphicx}

\begin{document}


\title{Relevance of the EEL spectroscopy for in-situ studies of the growth mechanism
of copper-phthalocyanine molecules on metal surfaces: Al(100)}


\author{A. Ruocco}
  \email{ruocco@fis.uniroma3.it}
\author{M.P. Donzello}
  \altaffiliation{present address: Dipartimento di Chimica, Universit\`{a} degli Sudi di Roma La Sapienza,
 P.zzale A. Moro 5 00185 Roma, Italy}
\author{F. Evangelista}
\author{G. Stefani}
\affiliation{Unit\`{a} INFM and Dipartimento di Fisica,
Universit\`{a} Roma Tre \\ Via della Vasca Navale 84, I-00146 Roma
Italia}


\date{\today}

\begin{abstract}
Reflection electron energy loss spectroscopy (EELS) in specular
and off specular geometry has been employed to study the early
stage of the copper phthalocyanine (CuPc) growth on Al (100)
substrate. EEL spectroscopy has been a useful tool in order to
study electronic structure of molecular films also in the
submonolayer regime. The electronic structure of the first
deposited layer of CuPc is strongly influenced by  charge transfer
from the Al substrate to the lowest unoccupied molecular orbital
(LUMO). The strong molecule-substrate interaction gives rise to a
coverage dependent frequency shift of the Al surface plasmon.
Successive layers have essentially the electronic structure of the
molecular solid. Momentum resolved EELS measurements reveal that,
in the case of the thicker film investigated (22 \AA), the plane
of the molecule is almost perpendicular to the surface of the
substrate.
\end{abstract}

\pacs{73.22.-f, 73.20.Mf, 79.20.Uv}

\maketitle


\section{Introduction}

Metal phthalocyanines, denoted MPc (Pc = phtalocyanine
C$_{32}$H$_{16}$N$_{8}$), have been synthesized using elements
from any group of the periodic table. They are planar molecules,
closely related to biological molecules such as porphyrins,
constituted by a porphyrazin ring (porphyrin-like) bonded to four
benzene rings. Main features of this class of molecules are a
metal atom in the center (usually one of the first transition
series) and an extended $\pi$-electron delocalization. MPc exhibit
an high chemical and thermal stability and exist in different
forms; among the various polymorphs \cite{Leznoff89}, the $\alpha$
and $\beta$ ones are the best known and the most widely studied.
In both crystalline forms the phthalocyanine units are positioned
in columnar stacks with the ring tilted with respect to the
stacking axis (tilt angle), which cohere to form the molecular
crystal. The two forms display identical interplanar distance (3.4
\AA) consistent with a Van der Waals bond, but they differ for the
tilt angle: $26.5^{\circ}$ in the $\alpha$ form and $45.8^{\circ}$
in the $\beta$ form. Furthermore their lattice parameters are
different (23.9 \AA  $\,$ in the $\alpha$ form and 19.6 \AA $\,$
in the $\beta$ form) as well metal-metal distance. The different
aggregation geometry results in changes of the electrical
conductivity along the stacking direction. From an electronic
point of view, MPc are semiconductors whose gap width depends on
the central atom (typically 1.5- 1.8 eV) and slightly on the
geometrical arrangement (a variation of 60-70 meV between the
$\alpha$ and $\beta$ form in the absorption spectra has been
observed \cite{Lucia68}). After p-doping by oxidizing agents and
stabilization of a face to face stacking \cite{Canadell84}, MPc
become electrical conductors thus enlarging their field of
applications. All these properties allow several technological
applications in different fields such as non-linear optics,
molecular electronics and highly specific gas-sensors fabrication
(such as NO$_{2}$ \cite{Dogo92a,Dogo92b}). Therefore the knowledge
of MPc spatial and electronic structure, both as bulk crystals and
as adsorbates on well-defined substrates, is much relevant. Over
the past decade MPc overlayers have been studied interfaced with
metals \cite{Tokito94,Smolyaninov00,Dufour94,Walzer01}
semiconductors \cite{Dufour94,Dufour95,Cox99a,Cox99b} and layered
compounds \cite{Ottaviano97,Shimada98,Walzer01}. Attempts have
been made to grow them in an ordered manner (heteroepitaxy) on
various substrates, some of them of high technological interest
\cite{Koma92,Tada92}. Substrate materials which have been shown to
support the epitaxial growth of phthalocyanine ultrathin films now
include single crystal metals, layered semiconductors (SnS$_{2}$,
MoS$_{2}$, HOPG), surface-passivated three dimensional
semiconductors (Si, GaAs, GaP), and insulators such as freshly
cleaved single crystal halide salts \cite{Leznoff96}.  In some
cases an overgrowth evolution characterized by different molecular
orientations has been observed. In particular, the adsorption
starts with planar arrangement at low coverages, when
substrate-molecule interaction is believed  to dominate, to change
into out-of-plane orientation at higher coverages where
molecule-molecule interaction is expected to become relevant
\cite{Poirier96}.

Recently, substrate surface reconstruction has been observed
induced by molecular deposition and it has been speculated that it
might be determined by strong molecule-substrate interaction. In
particular, for CuPc/Ag(110) interface, the Ag surface
reconstruction  has been ascribed to the presence of facets
\cite{Bohringer97}. More recently, for gold (110) surface has been
detected a clear long range reordering of the substrate due to
adsorption of CuPc \cite{Evangelista03}.
  The capability of $\pi$-conjugated molecules to induce displacements of surface atoms in conjunction
with self-organised molecular growth \cite{Yokoyama01} appears
then to be a general characteristic that is relevant from a
technological point of view and makes adsorption of this class of
molecules, in perspective, an attractive option for nano
manipulation of surfaces  \cite{Joachim00}.

Among  metals aluminum represents the archetypal for
nearly-free-electron systems; therefore the Al surface constitutes
an interesting alternative to the more extensively investigated
transition metal substrates. In particular the absence of occupied
$d$ orbitals might highlight the role played by localization of
the states in influencing the electronic structure of an
organic-inorganic interfaces. To the best of our knowledge Al
surface has never been studied  as substrate for CuPc films
deposition, apart from  the case in which it was used to support
thick films of CuPc. In this latter case, the aim was to analyze
bulk interband molecular transitions \cite{Scrocco93} and the
influence from the metal-molecule interface was not investigated.

The aim of this work is to study  the growth mode and the
molecular orientation of the overgrown film as a function of its
thickness, and to characterize  the geometric and electronic
structure of monolayer and submonolayer CuPc films, deposited by
thermal evaporation on an Al(100) substrate. CuPc is a dye pigment
thermally very stable, thus allowing purification by sublimation
technique and deposition by thermal evaporation or organic
molecular beam epitaxy (OMBE) \cite{Leznoff96}. Electron energy
loss spectroscopy (EELS) performed both  in reflection geometry
and as a function of the ejection angle are the main spectroscopic
tools applied in this work.

The literature reports on a limited number of  EELS experiments on
MPc films in general and on CuPc films in particular, most of them
made in transmission geometry on thick films, with high primary
electron energy. Low energy electron energy loss spectra in
reflection conditions have been measured on policrystalline films
of H$_2$Pc, CuPc, VOPc and PbPc \cite{Tada89,Bubnov74}, (prepared
by sublimation under high vacuum $10^{-7} \div 10^{-8}$ mbar) on a
Si substrate, with a primary electron energy of 100 eV. The
spectra of all phthalocyanine complexes examined have similar
structures and the observed peaks in the $\pi\rightarrow\pi^{*}$
excitations region are always in good agreement with the
correspondent optical absorption data \cite{Lucia68,Schechtman70}.

Although a limited number of works have made use of  EEL
spectroscopy to study MPc films, this technique is particularly
attractive because it allows to examine an energy range
corresponding to a region ranging from IR to soft X-ray in the
electromagnetic spectrum by using a laboratory based spectrometer
that can be easily conjugated to a growth chamber for in-situ
investigation of the growth mechanism. Main target of this paper
is to put in evidence the possibility to monitor the growth of
thin films of CuPc, ranging from submonolayer to  a few
monolayers, by the use of EEL spectroscopy and emphasize the
potentiality of EELS technique in measuring  the electronic
properties of these films. Furthermore, the possibility to
determine the orientation of the molecule with respect to the
substrate has been shown. The latter result has been reached
exploiting the relative orientation of transition dipole moment
(i.e. the symmetry of the  $\pi\rightarrow\pi^{*}$ transition at
3.7 eV) and momentum transferred in the collision. A similar
experiment, based on core transition, has already been performed
in the case of simpler organic molecule adsorbed on metallic
substrate \cite{Hitchcock89}.

\section{Experimental}

The experiments reported in this paper have been performed at the
LASEC laboratory (Dip. di Fisica and Unit\`{a} INFM,
Universit\`{a} Roma Tre) with an apparatus that allows to study
thin films, grown in-situ, by a variety of electron
spectroscopies, thus  providing complementary informations on both
electronic and geometric structure of the overlayer. In
particular, the apparatus consists of two separate UHV chambers.
The experimental chamber, equipped with an electron gun, a x-ray
source and two emispherical analyzers, is devoted to spectroscopic
investigations. A 5 degrees of freedom sample manipulator allows
to control position in space and temperature of the sample. A
comprehensive description of this apparatus is given elsewhere
\cite{Ruocco99a}.

The preparation chamber features an electron bombardment
evaporator Tricon \cite{Verucchi00}  and a quartz crystal
microbalance (QCM) used to control the growth rate of the films.
The specific nature of CuPc, i.e. the high condensability of the
evaporated  and its tendency to sublimate forming needle-shaped
crystals, made necessary to modify the evaporator source
\cite{Donzello00}. The thickness of the molecular overlayer is not
univocally determined by the rate of deposition onto the QCM as it
depends upon the sticking coefficient and the deposition mode.
Hence from the QCM measurements, a nominal thickness,
corresponding to uniform coverage of the surface, is deduced.
Furthermore, the preparation chamber features an ion gun for
sputtering substrates and a magnetically coupled linear
feedthrough for transferring samples.

The Al substrate is a single crystal (5 x 5 x 2.5 mm$^{3}$) with
(100) orientation and is supported by a molibdenum sample holder.
Two different cleaning procedures have been adopted  to remove the
thick oxide layer always present on samples stored in air. The Al
surface was cleaned by electro polishing prior to introduction in
the preparation chamber.  The electropolishing was performed on an
AB Electropolishing cell, Buchler Std, using a solution
constituted of 345 mL of HClO$_{4}$ 60 \% and 655 mL of
(CH$_3$CO)$_2$O. The sample was then cleaned under vacuum by
repeated cycles of sputtering with argon ions (4 keV, 6÷7 $\mu$A)
and annealing (450 $^o$C) \cite{Musket82}. The sample cleanliness
and order was checked before every deposition by means of  AES and
EELS . Commercial CuPc was obtained from Aldrich Chemical (97 \%
dye content); it was purified by sublimation under vacuum (540-550
$^o$C, $10^{-2} - 10^{-3}$ mbar) and then introduced in the
molibdenum crucible of the electron bombardment evaporator. The
purity of the powder was checked by elemental analyses for C, H,
N, performed by an EA 1110 CHNS-O CE instrument. Calculated for
C$_{32}$H$_{16}$CuN$_8$: C, 66.72; H, 2.80; N, 19.45. Found: C,
66.58; H, 2.66; N, 19.34 \%. (The uncertainty was $\pm$ 0.3 for C
and $\pm $ 0.1 for H and N). The molecule was sublimated onto the
substrate at room temperature at a rate of approximately 0.5
\AA/min. The film, prepared in such a way, has been found stable
until 400 $^o$C and under electron bombardment (impinging current
of few nA) does not suffer evident radiation damage; in summary it
stays clean, under UHV condition, for at least 48 hours. The EEL
spectra reported in the paper were collected at room temperature
with one of the two emispherical electron analyzers present in the
experimental chamber. EELS measurements were performed at  fixed
incident kinetic energy and the overall energy resolution was 500
meV throughout the whole range of incident electron energies (140
eV to 500 eV); the angular resolution was $\pm 0.5^{\circ}$.
 Two different kinds of EEL spectra have been
recorded in this work. In the first one the energy loss
probability is measured in specular reflection conditions with a
fixed incident angle of  34$^{\circ}$ from surface normal. In the
second one the probability for a given energy loss was measured as
a function of the transferred momentum  by rotating the sample
while keeping fixed the included angle between the incoming and
scattered beams.

\section{Results and discussion}

\begin{figure}
\includegraphics[bb=90 300 450 780,clip=true,width=9cm]{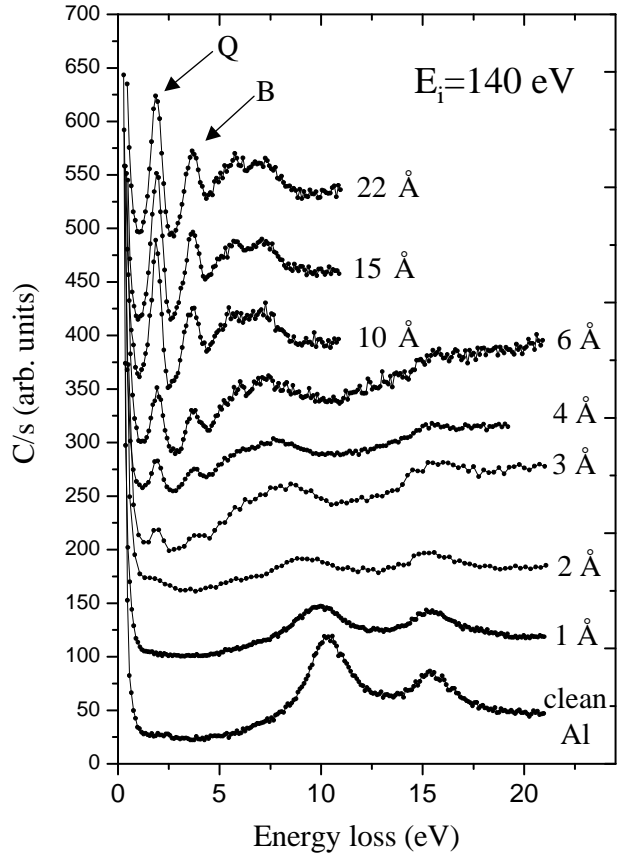}
\caption{\label{eels140}EEL spectra as a function of CuPc film
thickness performed in specular geometry with a primary energy of
140 eV. As a reference is also reported the EEL spectrum of the
clean Aluminum.}
\end{figure}

 In fig. \ref{eels140} are reported the constant
momentum transfer EEL spectra measured in specular reflection as a
function of CuPc coverage, the nominal thickness (from now on
coverage) ranges from 1 \AA $\,$ up to 22 \AA. All the spectra
have been collected in specular reflection geometry and with a
primary energy of 140 eV in order to take advantage of the reduced
mean free path  and then to be sensitive to the molecular film.
The EEL spectrum of clean Aluminum is also reported for reference.

\subsection{Interface plasmon}

\begin{figure}
\includegraphics[bb=150 50 480 510,clip=true,width=8.5cm]{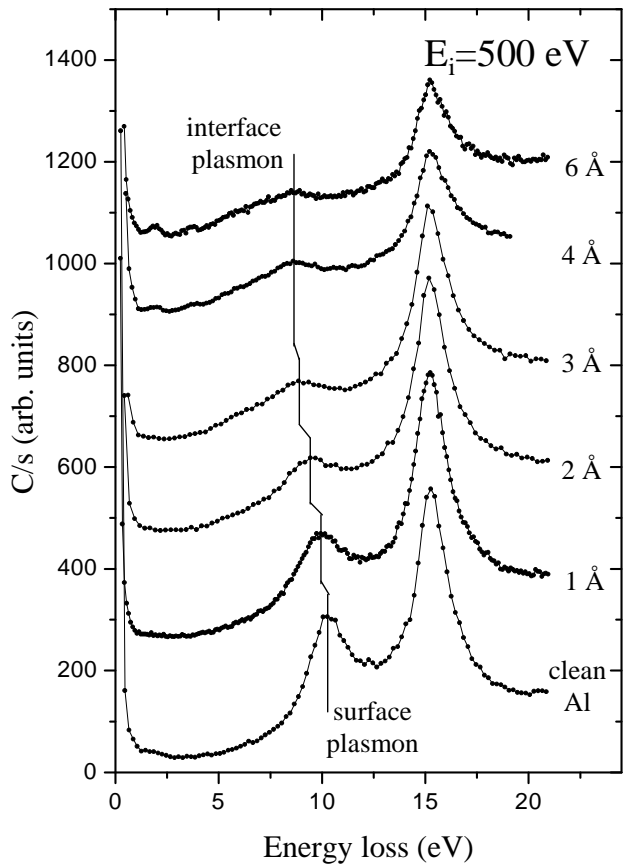}
\caption{\label{eels500}EEL spectra as a function of CuPc film
thickness performed in specular geometry with a primary energy of
500 eV. As a reference is also reported the EEL spectrum of the
clean Aluminum.}
\end{figure}

The clean Al spectrum is dominated by two structures at 10.5 eV
and 15 eV that correspond to the surface (SP) and bulk plasmon
(BP), respectively. The intensity of the bulk plasmon drops with
increasing coverage and  almost disappears already for a coverage
of 6 \AA. On the contrary  the surface plasmon shows a more
articulate  evolution. For the lowest coverages the centroid of
this structure shifts towards lower loss energy; the shift
increases as the thickness grows up to  3 \AA: in this situation
it is not any more possible to resolve the surface plasmon from
the molecular transitions appearing in the  5-8 eV region with
similar intensity. The evolution of the SP with coverage is more
evident in fig. \ref{eels500}, where are reported the EEL spectra
on the CuPc/Al(100) system, but with a primary energy of 500 eV,
in order to highlight the interface excitations rather than the
overlayer ones.
It is then easier to follow the evolution of surface plasmon at
very low coverages finding that it shifts  toward lower loss
energy when increasing the coverage and it reaches a steady value
(8.5 eV) already at 4 \AA, well before bulk-like conditions are
fulfilled. It is therefore plausible to ascribe the SP peak to an
interface plasmon such as those observed by Raether
\cite{Raether80} in the case of thin films deposited on metallic
substrate. More recently, a peak at 7 eV in the EEL spectrum of an
Al surface exposed to O$_{2}$ has also been attributed to  plasma
oscillations localized in the metal substrate \cite{Hoffman88}. To
ascertain whether or not the observed SP transition corresponds to
an interface plasmon, it can be considered that the energy of such
a collective mode  is expected to disperse with the overlayer
thickness unless  the following condition is  fulfilled
\cite{Raether80}:

\begin{equation}
q_sd \gg 1
 \label{eq1}
\end{equation}

where $q_s$ is the momentum associated with the plasma
oscillation, i.e. the surface component of the momentum exchanged
in the inelastic scattering  (q), and $d$ is the thickness of the
film. Dispersion of the plasma frequency as a function of the
thickness of the overlayer, was already observed for a thick Al
film (150 \AA) covered with oxide layer \cite{Kloos68}. In our
case $q_s$ can be evaluated on the basis of the double collision
model that is known to be valid
\cite{Ruocco99a,Ruocco99b,Saldin88} for the energy loss processes
in specular reflection geometry. According to this model, the
inelastic scattering is followed or preceded by an elastic one and
the inelastic cross section can be assumed different from zero
only for scattering angle falling within a cone of aperture
$\Theta_o = \Delta E/2E_o$ where $E_o$ is the energy of primary
beam and $\Delta E$ is the energy lost in the collision.
Consequently the maximum $q_s$ allowed is of the order of
$\Theta_o k_i \simeq 1.2 $ \AA $^{-1}$ where $k_i$ is the incident
momentum. This explains the dispersion of the interface plasmon
observed in fig.\ref{eels500}, in particular the asymptotic value
is reached for a coverage of 4 \AA $\,$ that is the minimum
coverage to satisfy inequality \ref{eq1} ($q_sd$ = 4.8).
 Moreover, making the assumption that the Al conduction electrons are well described by a free
electron gas, the interface plasmon frequency is related to the
dielectric constant of the molecular film by the relation
\cite{Raether80}:

\begin{equation}
\hbar\omega_s = \frac{\hbar\omega_p}{\sqrt{1+\varepsilon}}
 \label{eq2}
\end{equation}

From the measurements reported in fig.\ref{eels500} we estimate
for $\omega_s = 8.5$ eV and  for $\omega_p = 15$ eV hence
obtaining a value of 2.1 for the dielectric constant of the CuPc,
a value that well agrees with those measured for other planar
organic molecules, with an extended delocalization of $\pi$
electrons, like benzene. All these finding support the hypothesis
that the dispersing structure observed in fig.\ref{eels500} is to
be ascribed to a surface plasma wave propagating within the
aluminum substrate whose frequency is modified by the dielectric
response of the molecular adlayer. To the best of our knowledge
this is the first observation of an interface plasmon induced by
organic molecule in aluminum.

\subsection{Molecular transition}

Electronic transitions due to the CuPc molecule become evident in
the EEL spectrum starting from a coverage of 3 \AA; for this
coverage two weak peaks centered at 1.9 and 3.7 eV (hereafter Q
and B transitions respectively) appear together with a broad
structure between 5 and 8 eV that is more intense than the Q and B
transitions. Increasing the coverage, the Q and B transitions are
always present: their intensities increase as a function of the
coverage while their shape,  energy position and relative
intensity remain substantially unchanged. On the contrary the
broad structure at  5-8 eV shows a modest evolution reaching its
final shape at  6 \AA. For coverages of this value and higher the
structure shows two prominent features located at 5.8 and 7.1 eV
that are weaker than the B and Q transitions. It is interesting to
note that starting from 10 \AA, the EEL spectrum does not show
significant modifications, thus suggesting that the molecular film
has reached a bulk-like configuration. This is confirmed by the
observation that  the energies at which  electronic transitions
appear do correspond to those reported in a previous work on thick
CuPc films \cite{Tada89}. According to the diagram level of the
CuPc molecule as obtained from the four-orbital model
\cite{Schaffer72,Schaffer73,Gouterman78}, the transitions at lower
energies ($\Delta E < 5 $ eV) appearing in fig.\ref{eels140} are
assigned mostly to $\pi\rightarrow\pi^{*}$ electronic transitions
of phthalocyanine molecules. In particular the peak at 1.9 eV, the
Q band, is due to a mixture of  $a_{1u}(\pi) \rightarrow
e_g(\pi^*)$ and $b_{2g} \rightarrow b_{1g}$ transitions; although
the two transitions are almost degenerate in energy, the former
has a dipole moment perpendicular to the molecular plane while the
latter, mostly due to $d$ orbitals from copper atom, has dipole
moment in the plane of the molecule. On the contrary  the peak at
3.7 eV, the B band, is related to the single
$a_{2u}(\pi)\rightarrow e_g(\pi^*)$ transition thus having a well
defined symmetry with respect to the plane of the molecule. We
also observe peaks at 5.8 eV,  C band, and at 7.1 eV, $X_1$ band,
both assigned to a $\pi\rightarrow\pi^{*}$ transition. It is worth
noting that there is a good agreement between the transition
observed by means of EELS for the bulk-like coverages (10-22 \AA)
and those obtained in the optical absorption spectra for thick
film of CuPc \cite{Schechtman70}. Additionally, in absorption
spectroscopy two other bands at 4.7 eV (N) and at 7.8 eV ($X_2$)
have been identified. We speculate that the former transition (N)
gives rise, in our spectra, to the low energy shoulder of the peak
centered at 5.8 eV while the latter transition ($X_2$) is not
detectable as already reported in a previous  EELS work
\cite{Tada89}.

\begin{figure}
\includegraphics[bb=105 65 435 480,clip=true,width=8.5cm]{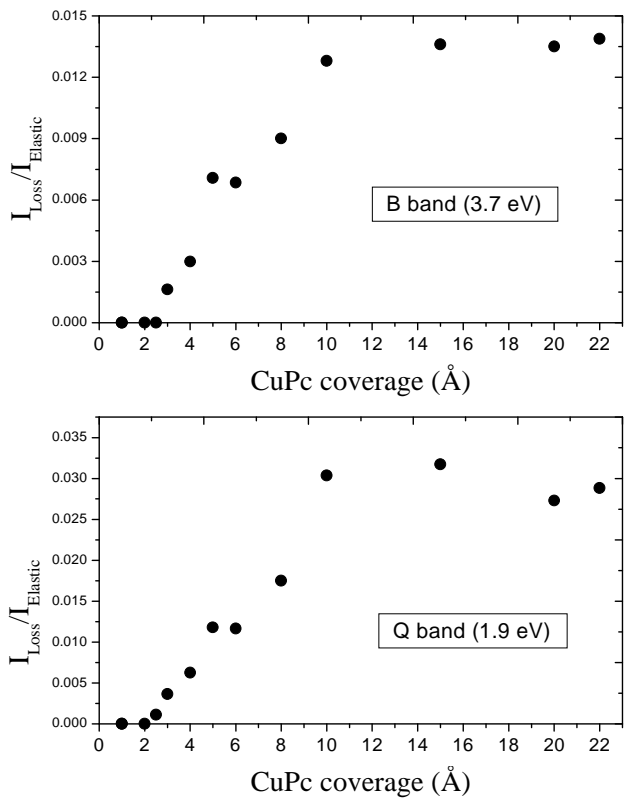}
\caption{\label{intqb}Q and B band intensity, normalized to the
related elastic peak intensity as a function of CuPc coverage.}
\end{figure}

From fig. \ref{eels140}, the Q and B transitions appear for
coverages of 3 \AA $\,$ and higher. This threshold value
correspond, in the hypothesis of a flat lying adsorption geometry,
to saturating the surface with one monolayer. In this framework
molecules from the first layer do not contribute to Q and B bands.
In order to understand whether at low coverages these structures
are simply confused in the background, or an alteration of the
electronic structure occurs that forbids them, the Q and the B
band intensities  are plotted as a function of CuPc coverage (see
fig. \ref{intqb}).
For coverages below 10 \AA $\,$ we observe that the experimental
data can be fitted with a straight line whose intercept to zero
corresponds to a coverage between 2 and 3 \AA. Hence, appearance
of the optical absorption bands (Q and B) in the EEL spectrum has
a clear threshold at a low non zero coverage. It implies that the
electronic structure of the first adsorbed molecules is different
from that of the bulk ones. We also notice that  the threshold
coverage of 2-3 \AA $\,$ is consistent with the saturation value
for one layer of flat lying absorbed molecules. This behavior is
confirmed by a similar investigation performed at 500 eV of
incident energy \cite{Donzello00}.  In other words we can make a
clear distinction between molecules directly bonded to the Al
substrate (coverage below 3 \AA) and molecules not directly bonded
to the substrate (coverage above 3 \AA). In the latter case the
electronic structure, as revealed by EELS, is identical to that of
bulk CuPc while in the former case modification of the electronic
structure is such to prevent transitions toward  the LUMO orbital.
Above  10 \AA $\,$ we observe a saturation of the Q and B bands
intensity. It is now important to understand why the optical
transitions are inhibited  for molecules directly bonded to the Al
substrate. The simplest hypothesis to be made requires that, as a
consequence of charge transfer,  electrons from the Al substrate
fill up  the molecular   $\pi^*$ LUMO and the $3d_{x^2-y^2}$
orbitals. In order to fill the LUMO (doubly degenerate) and the
$3d_{x^2-y^2}$ states, 5 electron per molecule should migrate from
the substrate. Roughly speaking, each CuPc molecule covers about
32 Al atoms, then each of this metallic atoms will contribute with
about 0.16 electron to the charge transfer process. Considering
the high density of nearly free electrons of the substrate, such a
charge transfer  is not unreasonable. Besides, a comparably large
charge transfer it has already been observed in the case of
C$_{60}$ overlayer grown on Al \cite{Hebard94}.
 Further support to the charge transfer mechanism comes from considerations on the molecular
energy levels. For molecular solids grown on solid surfaces it was
commonly assumed that organic- metal interface energy diagram can
be obtained by aligning the vacuum levels of the two materials.
Recently it has been demonstrated that this assumption is not
always true for both metal \cite{Hill98} and semiconductor
substrates \cite{Hill99}. Vacuum level alignment applies only when
the interaction between molecular film and substrate is weak. As
previously pointed out this is not our case. The charge transfer
with consequent formation of an ionic bond between CuPc and Al
substrate implies the formation of a dipole barrier at the
interface \cite{Hill99}. The presence of a surface dipole barrier
is also supported by the  consideration that the work function
(WF) of aluminum is 4.3 eV while the electron affinity (EA) of the
molecule (distance between LUMO and  vacuum level) is only 3.1 eV.
Then the 1.2 eV difference between the two levels should prevent
any charge transfer in the case of non-interacting interfaces. Our
EELS analysis suggests that LUMO is filled and then it is either
aligned or it lies below the Fermi level; we then conclude that
the LUMO state of the molecules directly bonded to the metal
shifts at least by 1.2 eV. Similar results have been already
observed for C$_{60}$/Au(110) \cite{Maxwell94}; also in that case
the authors claim for a charge transfer from the metal to the
molecule even though, in that case, the difference between the EA
and WF is  as large as 2.67 eV. In conclusion a sizeable charge
transfer from substrate to molecule explains the observed
threshold in coverage for appearance of optical bands in the EEL
spectra.

\subsection{EEL spectra as a function of the exchanged momentum}

The inelastic scattering in specular reflection geometry can be
described by means of the double collision model (DCM)
theoretically predicted \cite{Saldin88} and experimentally
verified \cite{Ruocco99a,Ruocco99b} for energy losses in the range
10$-$30 eV. In this model the measured electrons suffer a double
collision (one elastic and one inelastic) with the solid: in the
elastic collision electrons are  reflected from the surface, and
in the inelastic one an energy loss occurs while scattering in
forward direction. Concerning our experiment, the proposed model
has two main implications: i) the smallness of the momentum
exchanged in the inelastic collision, compared with the momentum
of the incident electrons, allows to apply the dipole
approximation, ii) two channels will incoherently contribute to
the total cross section, that in which the elastic collision
precedes (D+L) and that in which it follows (L+D) the inelastic
collision. As previously pointed out the transitions at the lowest
energies have a well defined symmetry; in particular the B band,
due to its unambiguous   $\pi\rightarrow\pi^{*}$ character, has
dipole moment ($\bf p$) perpendicular to the plane of  the
molecule. The spatial orientation of the molecule can then be
probed by changing, in the experiment,  the direction of dipole
moment with respect to the momentum transferred in the inelastic
collision. In this framework the cross section of EELS is
proportional to ${\bf |p \cdot q|}^2$, thus  the molecule
orientation can be derived studying the EELS probability of the
transition B as a function of the angle included between $\bf p$
and $\bf q$.
\begin{figure}
\includegraphics[bb=140 130 495 400,clip=true,width=8.5cm]{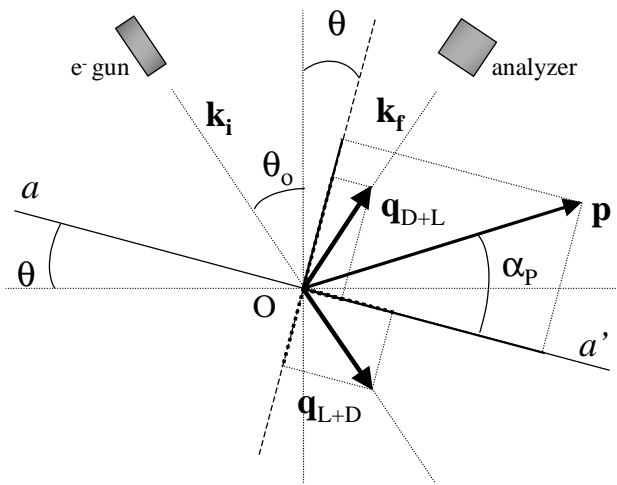}
\caption{\label{geometry}Experimental geometry for the EEL
spectrum as a function of the exchanged momentum. Electron gun and
the analyzer are fixed; $\alpha_p$ is  the angle between the
dipole moment of the molecular transition and the surface of the
sample described by the plane \textit{aa'}. The rotation of the
surface is described by the angle $\theta$ and $\theta_{\circ}$ is
the angle between the electron gun (the analyzer) and the normal
to the surface when specular conditions are satisfied.}
\end{figure}
In fig.\ref{geometry} is reported the kinematics  of the
experiment. According to the DCM, two inelastic exchanged momenta
are drawn directed along the directions of the incoming beam
(${\bf q}_{L+D}$) and of the collected beam (${\bf q}_{D+L}$).
$\alpha_P$ is the angle formed by vector $\bf p$ with respect to
the surface and it represents the orientation of the molecule. The
experiment was performed with the higher coverage (22 \AA), by
scanning the  angle $\theta$, i.e. the angle between the normal to
the surface and the bisector of the included angle between
incoming and outgoing  electron beams (see fig.\ref{geometry}).
The directions of incident and diffracted beams are fixed in the
laboratory reference frame, thus also (${\bf q}_{D+L}$) and (${\bf
q}_{L+D}$) have a fixed direction. By keeping fixed the energy of
incoming and outgoing electrons, the rotation of the sample
results in a rotation of ${\bf p}$ and thus in a variation of the
scalar product ${\bf p \cdot q}$ ($q$ is not univocally
determined, 2 values are present for each kinematics). It is more
convenient to describe the process in the sample reference frame,
where the scalar product between the exchanged momentum and dipole
moment can be separately written for the two channels (L+D and
D+L) as:

\begin{eqnarray}
{\bf q \cdot  p}_{L+D} = [q_{\perp} p_{\perp}+ {\bf
q_{\parallel}\cdot p_{\parallel}}]_{L+D}  \nonumber
\\
= qp \, [\cos(\theta_{\circ}-\theta) \sin \alpha_p +
\sin(\theta_{\circ}-\theta) \cos\alpha_p \cos \beta]\nonumber
\\ \nonumber
\\
{ \bf q \cdot p}_{D+L} = [q_{\perp} p_{\perp}+ {\bf
q_{\parallel}\cdot p_{\parallel}}]_{D+L}  \nonumber
\\
= qp \, [\cos(\theta_{\circ}+\theta) \sin \alpha_p +
\sin(\theta_{\circ}+\theta) \cos\alpha_p \cos \beta] \label{eq3}
\end{eqnarray}

where the subscripts L+D and D+L take in account the presence of
the two possible scattering channels previously discussed,$\beta$
represents the orientation of the molecule in the azimuthal plane
and for  $\beta =0$,${\bf p}$  lies in the scattering plane. The
differential inelastic cross section will then be, as already
shown elsewhere \cite{Ruocco99a,Ruocco99b},  the incoherent sum of
the D+L and L+D cross sections. Hence, within First Born dipolar
approximation the energy loss differential cross section factors
out in a kinematical term times the optical oscillator strength of
the transition involved, times the sum of the two orientation
terms appearing in \ref{eq3}. Taking into account that none of the
studied interfaces has displayed a LEED pattern, an azimuthal
random orientation of the adsorbed molecules can be safely
assumed. This being the case, the dependence of the inelastic
cross-section upon the polar angle reduces to the modulus square
of the orientation factors averaged over the azimuthal angle beta,
which is

\begin{eqnarray}
\left| {\bf q \cdot p} \right|^2 \propto R \cdot
[\cos^2(\theta_{\circ}-\theta) \sin^2 \alpha_p  \nonumber
\\
+\frac{1}{2} \sin^2(\theta_{\circ}-\theta) \cos^2\alpha_p]_{L+D}
\nonumber
\\
+ (1-R) [\cos(\theta_{\circ}+\theta) \sin^2 \alpha_p  \nonumber
\\
+\frac{1}{2} \sin^2(\theta_{\circ}+\theta) \cos^2\alpha_p]_{D+L}
 \label{eq4}
\end{eqnarray}

R is the relative weight of the two channels and it depends
essentially on the amplitude of the elastic component of the cross
section \cite{Ruocco99a}.
\begin{figure}
\includegraphics[bb=120 140 555 465,clip=true,width=8.5cm]{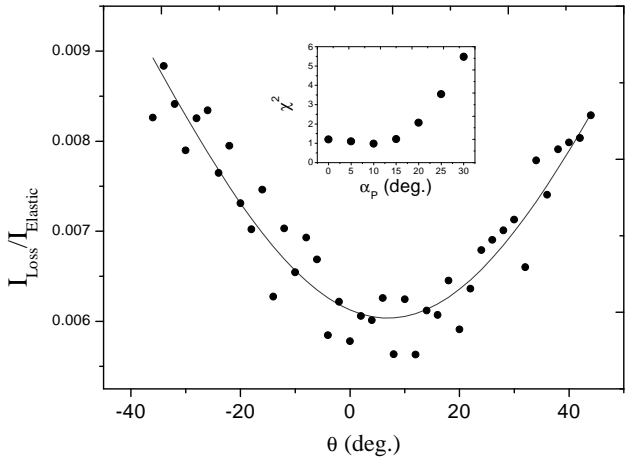}
\caption{\label{angres}B band intensity ($\bullet$), normalized to
the related elastic peak intensity, as a function of the polar
angle. The continuous line is the best fit obtained with the trial
function (\ref{eq4}) and $\alpha_p = 15^{\circ}$. In the insert
$\chi^2$ values as a function of the $\alpha_p$ parameter are
reported.}
\end{figure}
In fig.\ref{angres} is reported the intensity of the transition B,
normalized to the related elastic peak intensity, as a function of
the polar angle, expression \ref{eq4} is a trial function with the
scaling factor R used as a free parameter. The fitting procedure
has been repeated for several value of $\alpha_p$ ranging from
zero to 90$^\circ$. In the insert of fig.\ref{angres} is reported
the $\chi ^2$ as a function of $\alpha_p$ from which it is
possible to conclude that the proposed model can be accepted for
values of $\alpha_p$ in the range 0-15$^\circ$ where the $\chi ^2$
is almost constant and equal to 1. From the structural point of
view this correspond to have the plane of the molecule oriented
almost perpendicular to the surface plane. The best R value  is
0.54 that suggests an almost equal probability for L+D and D+L
scattering channels. This finding is in agreement with previous
similar experiments on clean  surfaces \cite{Ruocco99a} that
support the hypothesis of incoherent superposition of the two,
equally relevant, scattering channels contributing to the EEL
spectrum. Once more, momentum resolved EELS from adsorbed
molecules has been shown to be a sensitive, accurate tool for
determining orientation with respect to the substrate surface of
thin molecular films.

\section{Conclusions}

In conclusion, the early stage of absorption of CuPc on the
Al(100) surface has been studied by angle resolved EELS. The
method has proven to be very sensitive and it allows to
investigate coverages as low as 1 \AA. The first monolayer of
molecules suffers a massive charge transfer from the substrate
 inhibiting the optical B and Q absorption bands that are
instead characteristic of bulk aggregation. This finding suggests
that CuPc interacts with the Al substrate via a strong ionic bond.
The molecule substrate interaction is also testified by the
observed shift in frequency of the Al surface plasmon. For the
thick overgrown film the molecular plane is oriented predominantly
perpendicular to the substrate plane and, in contrast to the first
adsorbed layer, it does not show evidences for charge transfer
from the substrate (B and Q band are restored).

\bibliography{biblio}



\end{document}